\documentclass[%
reprint, 
superscriptaddress,
nofootinbib,
 amsmath,amssymb,
 aps, prresearch,
]{revtex4-2}

\usepackage{xcolor}
\usepackage{soul}
\usepackage[normalem]{ulem}

\usepackage{siunitx}
\usepackage{enumitem}

\usepackage{graphicx}
\usepackage{dcolumn}
\usepackage{bm}

\usepackage[colorlinks=true, linkcolor=blue, citecolor=blue, urlcolor=blue]{hyperref}
\usepackage[capitalise]{cleveref}
\crefname{section}{Sec.}{Secs.}
\usepackage{mathtools}
\usepackage{physics}

\usepackage{float}
\makeatletter
\let\newfloat\newfloat@ltx
\makeatother

\usepackage{algorithm}
\usepackage[noend]{algpseudocode}
\usepackage{orcidlink}

\usepackage{amsthm}
\newtheorem{theorem}{Theorem}[section]
\newtheorem{definition}{Definition}[section]
\newtheorem{remark}{Remark}[section]

\newcommand{\squares}{s}

\DeclarePairedDelimiter\ceil{\lceil}{\rceil}

\def\F2{\mathbb{F}_2} 

\DeclareMathOperator{\Ker}{ker}
\DeclareMathOperator{\Ima}{im}

\makeatletter
\def\algbackskip{\hskip-\ALG@thistlm}
\makeatother

\makeatletter
\renewcommand{\ALG@name}{Algorithm }
\makeatother

\begin{document}

\preprint{APS/123-QED}

\title{\textbf{Almost linear decoder for optimal geometrically local quantum codes} 
}%

\author{Quinten Eggerickx\,\orcidlink{0000-0002-4709-3115}}
    \email{quinten.eggerickx@imec.be}
    \affiliation{Imec, Kapeldreef 75, 3001 Leuven, Belgium}
    \affiliation{Department of Electrical Engineering, KU Leuven, Kasteelpark Arenberg 10, 3001 Leuven, Belgium}
    \affiliation{Department of Physics and Astronomy, Ghent University, Krijgslaan 281, 9000 Gent, Belgium}
    
\author{Adam Wills\,\orcidlink{0009-0005-6653-8879}}
    \affiliation{Center for Theoretical Physics, Massachusetts Institute of Technology, Cambridge, Massachusetts 02139, USA}
    \affiliation{Hon Hai Research Institute, Taipei, Taiwan}
    
\author{Ting-Chun Lin\,\orcidlink{0000-0002-8994-4598}}
    \affiliation{Department of Physics, University of California San Diego, San Diego, California 92093, USA}
    \affiliation{Hon Hai Research Institute, Taipei, Taiwan}
    
\author{Kristiaan De Greve\,\orcidlink{0000-0002-1314-9715}}
    \affiliation{Imec, Kapeldreef 75, 3001 Leuven, Belgium}
    \affiliation{Department of Electrical Engineering, KU Leuven, Kasteelpark Arenberg 10, 3001 Leuven, Belgium}

\author{Min-Hsiu Hsieh\,\orcidlink{0000-0002-3396-8427}}
    \affiliation{Hon Hai Research Institute, Taipei, Taiwan}

\date{\today}

\begin{abstract}
    Geometrically local quantum codes, which are error correction codes embedded in $\mathbb{R}^D$ with checks acting only on qubits within a fixed spatial distance, have garnered significant interest. Recently, it has been demonstrated how to achieve geometrically local codes that maximize both the dimension and the distance, as well as the energy barrier of the code. In this work, we focus on the constructions involving subdivision and show that they have an almost linear time decoder, obtained by combining the decoder of the outer good qLDPC code and a generalized version of the Union-Find decoder. This provides the first decoder for an optimal geometrically local three-dimensional code. We demonstrate the existence of a finite threshold error rate under the code capacity noise model using a minimum weight perfect matching decoder. Furthermore, we argue that this threshold is also applicable to the decoder based on the generalized Union-Find algorithm.
\end{abstract}

\maketitle


\section{Introduction}

Large-scale fault-tolerant quantum computers hold the promise of revolutionizing our world by being able to solve certain computational problems far beyond the reach of conventional computers. A key element for fault-tolerance is the quantum error correction (QEC) code, which encodes the logical state of the system over many physical qubits and in that way protects the logical state against noise.

However, over time, errors accumulate and logical errors start to become more and more likely. Encoding on its own is therefore not enough. One also needs a decoder that is capable of predicting which errors occurred, such that these errors can be corrected. Moreover, it is crucial for fault-tolerant computation that this decoding process can be performed efficiently \cite{terhal2015quantum, skoric2023parallel}.

Among the different quantum codes, quantum low-density parity-check (qLDPC) codes have garnered significant attention. Practically, qLDPC codes are preferred because each stabilizer operates on only a limited number of qubits, and each qubit is operated on by only a limited number of stabilizers. This low density feature is advantageous for experiments, as it limits the spread of errors during the extraction of the syndrome. Theoretically, qLDPC codes have become the focus of research since Gottesman \cite{gottesman2013fault} showed that fault-tolerant computation can be achieved with only a constant overhead by using qLDPC codes with good properties. Furthermore, recent developments over the past few years have led to constructions \cite{panteleev2022asymptotically, leverrier2022quantum, dinur2023good} of \emph{good} qLDPC codes achieving asymptotically optimal linear dimension and distance.

Although good qLDPC codes accomplish the best performance possible, it is well-established \cite{bravyi2009no, bravyi2010tradeoffs} that these codes cannot be implemented solely with local interactions, which poses a significant challenge to their practical implementation. Recently, this issue was addressed in \cite{portnoy2023local, lin2023geometrically, williamson2023layer, li2024transform} by taking a good qLDPC code and transforming it into a geometrically local code. Evidently, the resulting code is no longer a \emph{good} code, but does saturate the Bravyi-Poulin-Terhal (BPT) bound \cite{bravyi2009no, bravyi2010tradeoffs} and achieves an optimal energy barrier.

In this paper, we construct a decoder for these geometrically local quantum codes.

\subsection{Main contribution}

Our main contribution is the construction of an almost linear-time serial decoder for the geometrically local codes from subdivision, developed in \cite{lin2023geometrically} and later generalized in \cite{li2024transform}, by extending the Union-Find decoder \cite{delfosse2021almost} to the generalized surface code. 

\begin{theorem}[informal]
    The decoder for the subdivided code, as described in \cref{alg:decoder_subdivided}, runs in almost linear time $O(n\alpha(n))$, where $\alpha(n)$ is the inverse Ackermann function, and can correct an error of weight up to a constant fraction of the distance.
\end{theorem}

This provides a positive answer to the first question raised in \cite{lin2023geometrically} by showing that indeed an efficient decoder, capable of correcting a number of errors up to a constant times the distance, can be obtained by combining the decoder for the surface code and the decoder for the original qLDPC code.

Moreover, our decoder can also be applied at once to the more general optimal geometrically local codes of \cite{li2024transform}. Although the local product structure is lacking in this generalized construction, all the elements that are needed for our decoder, i.e. patches of surface codes on top of a qLDPC scaffold, are present.

Our decoder is the first decoder for optimal geometrically local quantum codes in any dimension. In particular, it is the first almost linear time decoder for optimal three-dimensional (3D) geometrically local codes. While previous work has demonstrated fast serial and parallel decoding in three dimensions by combining two-dimensional decoding subproblems \cite{brown2020parallelized}, these approaches were not applied to optimal codes.

Although much research is being conducted on the 3D integration of superconducting qubits, experimental progress remains in its early stages. Current demonstrations are largely limited to flip-chip integration, where a quantum chip and a control chip are connected using through-silicon vias. In this layout, qubits reside on the quantum chip, while readout resonators and control lines are housed on the control chip \cite{kosen2022building,rosenberg20173d,yost2020solid}. While there is no experimental realization yet, these techniques could, in principle, be extended to stack multiple quantum chips, creating a 3D lattice of qubits. Moreover, advances in 3D integration for classical Complementary Metal-Oxide-Semiconductor (CMOS) technology \cite{zhang2022challenges} could inform and potentially accelerate the development of such architectures for superconducting qubits.

For spin qubits, the looped pipeline architecture represents a promising approach to achieving effective 3D local connectivity. While this concept has not yet been implemented at the scale envisioned (e.g., Fig. 4 in \cite{cai2023looped}), significant milestones have been reached. Coherent spin shuttling through a minimal loop of three germanium quantum dots has been demonstrated with an average fidelity per shuttle of $F \simeq 99.63\%$ \cite{van2024coherent}. This can be viewed as a foundational step towards building larger looped pipeline architectures. Similarly, recent achievements in silicon quantum dots include spin shuttling over an accumulated distance of \SI{10}{\micro\meter} (100 quantum dots) with an average fidelity per shuttle of $F \simeq 99.99\%$ \cite{de2024high}. Extrapolating these results to the looped pipeline architecture suggests that a 3D code with 100 layers in the third dimension could be feasible.

While these advances are promising, significant technical challenges remain before practical implementation of 3D geometrically local codes becomes a reality. Nonetheless, these developments highlight the potential for further exploration and innovation in this area.

Our linear time decoder thus supports the ongoing development of these systems, allowing them to scale to practically useful sizes.\\

We also considered the subdivided code under the code capacity noise model and showed that, using a combination of minimum weight perfect matching and a good qLDPC decoder, there exists a finite threshold error rate.

\begin{theorem}
    Geometrically local codes from subdivision, together with a combined minimum weight perfect matching and good qLDPC decoder, have a finite threshold error rate. Moreover, logical errors are suppressed exponentially with the subdivision parameter $L$.
\end{theorem}

This is a first demonstration of a nontrivial threshold for optimal geometrically local codes in any dimension, thereby strengthening the argument for their practical application.

Additionally, we present a reasoned argument suggesting that this result may also apply to the almost linear-time subdivided decoder outlined in \cref{alg:decoder_subdivided}.

\subsection{Construction and proof overview}

\begin{figure*}
  \centering
  \includegraphics[width=0.8\textwidth]{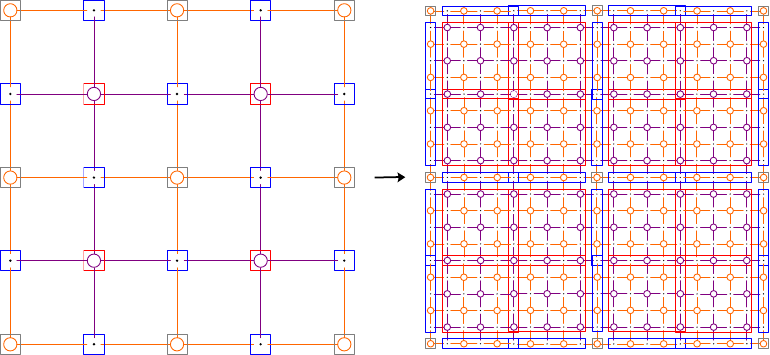}
  \caption{A scaffold code gets filled in with squares of the surface code. $Z$ stabilizers, qubits and $X$ stabilizers are represented by orange circles, black dots and purple circles, respectively. Three distinct types of regions can be identified in the subdivided code. The type 1 regions (patches of generalized surface code) are indicated in red and the type 2 regions (sections of generalized repetition code) are indicated in blue. The type 3 regions are not specially indicated as they simply correspond to the original $Z$ stabilizers.}
  \label{fig:multiple_patches}
\end{figure*}

\begin{figure}[b]
  \centering
  \includegraphics[width=\linewidth]{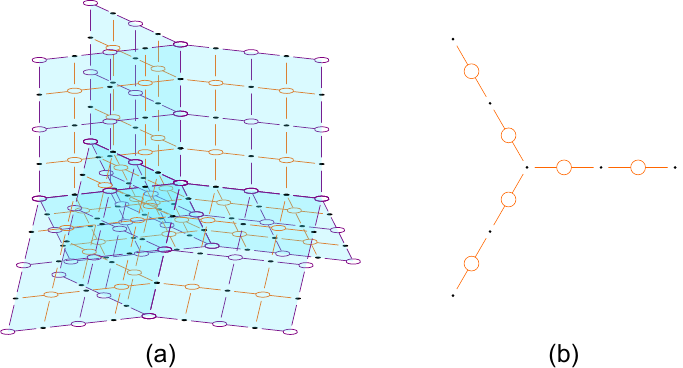}
  \caption{(a) Generalized surface code and (b) generalized repetition code.}
  \label{fig:gen_surf_and_rep_code}
\end{figure}

Since the geometrically local codes from subdivision are formed by taking a good qLDPC code as a scaffold and filling in the gaps with $L\times L$ squares of the surface code, one can identify three distinct types of regions in the code (\cref{fig:multiple_patches}). The first type are squares of the surface code surrounding an $X$ stabilizer in the original good qLDPC code. Due to the structure of the good qLDPC code, this region will form a patch of a generalized surface code where multiple squares of the surface code are glued together [\cref{fig:gen_surf_and_rep_code}(a)]. The second type of region are the boundaries of the patches of generalized surface code. Similarly, these form generalized repetition codes [\cref{fig:gen_surf_and_rep_code}(b)]. The last type of region are the boundaries of the type two region, and they correspond to the $Z$ stabilizers in the good qLDPC code.

The decoder of the subdivided code is constructed by applying a decoder to each of the three types of regions. The decoder on the patches of generalized surface code corrects errors to the best extent possible and pushes any remaining errors to its boundary. The decoder on the sections of generalized repetition code then takes on these additional errors and also corrects them as best as possible and pushes any remaining errors to its boundary. Finally, the decoder of the good qLDPC code takes care of the remaining errors.

The proof of the correctness and the time-complexity of the decoder primarily focuses on dealing with hyperedges and clusters of errors spanning over multiple squares of surface code. The key property that is repeatedly used to solve these issues is the low-density property of the underlying good qLDPC code.

At a high level, our approach relies on the fact that the subdivision code is chain homotopy equivalent to the original good qLDPC code, a strategy also discussed in Appendix A of Ref. \cite{hastings2021fiber}. Our decoding procedure essentially consists of two stages: the first maps the subdivided code back to the original code, and the second applies the decoder for the original qLDPC code. Both stages can be naturally extended to the codes in \cite{portnoy2023local,williamson2023layer}.

\subsection{Future directions}

\paragraph{Fast parallel decoder?}
There are several fast parallel decoders for the surface code under random noise \cite{fowler2013minimum, skoric2023parallel, wu2023fusion, das2022afs, tan2023scalable, liyanage2023scalable}. Can these be used to lead to a faster parallel decoder for the subdivided code under random noise?

\paragraph{Application of parallel window decoding?}
It was recently demonstrated \cite{skoric2023parallel,bombin2023modular} how to parallelize a decoder using parallel window decoding. Can the subdivided decoder, presented in this paper, be modified such that this procedure can be applied? Can it be applied in both the space and time directions?

\section{Preliminary}

\subsection{Chain complexes}

\begin{definition}[Chain complex]
    A chain complex $X$ consists of a sequence of vector spaces $\mathbb{F}_2^{X(i)}$ generated by sets $X(i)$, along with linear maps $\delta_i: \mathbb{F}_2^{X(i)} \to \mathbb{F}_2^{X(i+1)}$ known as coboundary operators, where the coboundary operators satisfy
    \begin{equation*}
        \delta_{i+1}\delta_i = 0\ .
    \end{equation*}
\end{definition}

Given a chain complex, one obtains the dual chain complex by reversing the direction of the linear maps. The coboundary operators $\delta_i$ then become boundary operators $\partial_i$. By using the elements of $X(i)$ as a canonical basis of $\mathbb{F}_2^{X(i)}$, the boundary operators $\partial_i: \mathbb{F}_2^{X(i)} \to \mathbb{F}_2^{X(i-1)}$ can be defined as the matrix transpose of $\delta_{i-1}$, $\partial_i\coloneqq \delta_{i-1}^{T}$. Note that the dual chain complex is also a chain complex since
\begin{equation*}
    \partial_{i-1}\partial_{i} = 0\ .
\end{equation*}

We provides some standard terminology. Elements of the kernel of the boundary operators are called cycles, while elements of the kernel of the co-boundary operators are called co-cycles
\begin{subequations}
    \begin{eqnarray*}
        Z_i &\coloneqq \Ker \partial_i = \{c_i \in \F2^{X(i)} : \partial_i c_i = 0\}\ ,\\
        Z^i &\coloneqq \Ker \delta^i = \{c^i \in \F2^{X(i)} : \delta^i c^i = 0\}\ .
    \end{eqnarray*}
\end{subequations}
Similarly, elements of the image of the boundary operators are called boundaries, and elements of the image of the co-boundary operator are called co-boundaries
\begin{subequations}
    \begin{eqnarray*}
        B_i &\coloneqq \Ima \partial_{i+1} = \{\partial_{i+1} c_{i+1} : c_{i+1} \in \F2^{X(i+1)}\}\ ,\\
        B^i &\coloneqq \Ima \delta^{i-1} = \{\delta^{i-1} c^{i-1} : c^{i-1} \in \F2^{X(i-1)}\}\ .
    \end{eqnarray*}
\end{subequations}
Since $\partial_i \partial_{i+1} = 0$ it follows that $B_i \subset Z_i$.
When $B_i = Z_i$ the chain complex is said to be \emph{exact} at $i$.


\subsection{Classical and quantum error-correcting codes}

\subsubsection{Classical error-correcting codes}

A classical binary code is a $k$-dimensional linear subspace $C \subset \F2^n$, which is defined as the kernel of a parity-check matrix $H: \F2^n \to \F2^m$. $n$ is called the size and $k$ is called the dimension of the code. The distance $d$ is the minimum Hamming weight of a nontrivial codeword
\begin{equation*}
    d = \min_{c \in C-\{0\}} \abs{c}\ .
\end{equation*}\\

The energy barrier $\mathcal{E}$ is defined as the minimum energy among all walks $\gamma$ from $0$ to a nontrivial codeword
\begin{equation*}
    \mathcal{E} = \min_{\gamma_{0\rightarrow c}, c \in C - \{0\}} \epsilon\left(\gamma_{0\rightarrow c}\right)\ ,
\end{equation*}
where a walk $\gamma_{0\rightarrow c}$ is a sequence of vectors $(C_0 = 0, c_1, \ldots , c_t = c)$ with $\abs{c_i - c_{i+1}} = 1$, and the energy of a walk is the maximum energy of the vectors in it: $\epsilon(\gamma) = \max_{c_i \in \gamma} \abs{H c_i}$.\\

Most often we are interested in the scaling behavior of codes. Therefore, we do not just consider one specific code, but rather a whole family of codes where the size of the code $n$ is taken as a parameter of the family. A code family is called a low-density parity-check (LDPC) code if each row and each column of the parity-check matrix has a bounded number of nonzero elements. A code family is said to be \emph{good} if both the dimension and the distance scale linearly with $n$: $k = \Theta(n)$ and $d = \Theta(n)$.


\subsubsection{Quantum error-correcting codes}

In general, a quantum code is defined to be a linear subspace of the Hilbert space of the appropriate dimension. For a system of $n$ qubits, this Hilbert space is $\mathbb{C}^{2^n}$. \\

An important class of quantum codes are the so-called \emph{stabilizer codes}. For these codes the codespace (subspace) is determined as the simultaneous $+1$ eigenspace of an Abelian subgroup of the $n$-fold Pauli group, called the \emph{stabilizer group} $\mathcal{S}$. Elements of $\mathcal{S}$ are called \emph{stabilizers}.

The correction of Pauli errors in a stabilizer code is based on the measurement of the stabilizers. The result of these measurements is called the \emph{syndrome} of the error. The task of the \emph{decoder} is then to take the syndrome as an input, and determine which correction needs to be applied to avoid a \emph{logical error}. Logical errors commute with all the stabilizers, but they are not stabilizers themselves. As such they act non-trivially on the codespace and must therefore be avoided.


\subsubsection{Quantum erasures}\label{sec:quantum_erasures}

Quantum erasures are a special type of error where we know which qubits have suffered an error, but we don't know what the error was. We say that some qubits have been erased. The set of erased qubits is called the \emph{erasure} and is denoted by $\varepsilon$.

For erasures, the decoding task is greatly simplified since we only need to find an error $\tilde{P}\subset \varepsilon$ that is consistent with the syndrome to find the most likely correction. Since operators of the stabilizer $S$ act trivially on the code space, the most likely correction is a coset $\tilde{P}S$ that maximizes the conditional probability $\mathbb{P}(PS|\varepsilon, \sigma)$.

\begin{theorem}[Lemma 1 of \cite{delfosse2020linear} (generalized to any stabilizer code)]\label{thm:}
Given an erasure $\varepsilon$ for a stabilizer code with stabilizer $S$ and a measured syndrome $\sigma$, any coset $\tilde{P}S$ of a Pauli error $\tilde{P} \in \varepsilon$ of syndrome $\sigma$ is a most likely coset.
\end{theorem}


\subsubsection{Quantum CSS codes}

A quantum CSS code \cite{calderbank1996good,steane1996error} is a stabilizer code, specified by two classical codes $C_z$, $C_x$ represented by their parity-check matrices $H_z: \F2^n \to \F2^{m_z}$, $H_x: \F2^n \to \F2^{m_x}$ which satisfy $H_x H_z^T = 0$. This condition allows us to associate a three-term chain complex to the quantum code,
\begin{equation}
    X: \mathbb{F}_2^{m_x}\xrightarrow{\delta_0 = H_x^T} \mathbb{F}_2^n \xrightarrow{\delta_1 = H_z} \mathbb{F}_2^{m_z}\ .
\end{equation}
Here, $n$ is the number of qubits, $m_x$ is the number of $X$ checks and $m_z$ is the number of $Z$ checks. The dimension $k = \dim C_x - \dim C_z$ is the number of logical qubits. The distance $d = \min(d_x,d_z)$ where
\begin{equation}
    d_z = \min_{c_z \in C_z - C_x^\perp} \abs{c_z}\ , \qquad
    d_x = \min_{c_x \in C_x - C_z^\perp} \abs{c_x}
\end{equation}
are the $Z$ and $X$ distance.


\subsection{Kitaev surface codes}

Kitaev surfaces codes \cite{kitaev2003fault,kitaev_surface_code_2024} are quantum CSS codes where the qubits are placed on a two-dimensional surface and where all parity checks are local, i.e. they only involve nearest-neighbor interactions. More specifically, given a surface and a cellulation of that surface described by $G = (V,E,F)$ with $V$ the set of vertices, $E$ the set of edges and $F$ the set of faces, the code is defined by placing a qubit on each edge $e \in E$, a $Z$-parity-check $Z(v)$ on each vertex $v \in V$, and an $X$-parity-check $X(f)$ on each face $f \in F$ \cite{delfosse2016generalized}.\\

Since $Z(v) = \prod_{v \in e} X_e$ only acts on the qubits of edges incident to $v$, and $X(f) = \prod_{e \in f} Z_e$ only acts on the qubits of edges on the boundary of $f$, the code is local. Moreover, the code is geometrically local for geometrically local tessellations (see \cref{def:geom_local}).\\

Note that both surfaces with and without boundaries can be considered .

\subsection{Subdivided codes}

\begin{definition}[Geometrically local in $D$ dimensions]\label{def:geom_local}
    A code family $\{C(n)\}$ is called geometrically local in $D$ dimensions if there exist a locality parameter $a \in \mathbb{R}_{>0}$, a density $b \in \mathbb{R}_{>0}$ and an embedding in $\mathbb{Z}^D$ such that for each $C(n)$
    \begin{enumerate}[label=(\roman*)]
        \item the Euclidean distance between each check and the qubit it interacts with is $\leq a$\ ,
        \item the number of qubits and checks located at each lattice point in $\mathbb{Z}^D$ is $\leq b$\ .
    \end{enumerate}
\end{definition}

Several good quantum LDPC codes have been identified \cite{panteleev2022asymptotically,leverrier2022quantum,dinur2023good}. Despite these codes exhibiting optimal scaling behavior, it is well established that they cannot be realized using only local interactions \cite{bravyi2009no}. Restricting oneself to geometrically local codes, the distance and dimension of the code are upper bounded by the BPT bound \cite{bravyi2009no,bravyi2010tradeoffs}: 
\begin{equation}\label{eq:BPT}
    d = O\left(n^{\frac{D-1}{D}}\right)\ , \qquad
    k = O\left(n^{\frac{D-2}{D}}\right)
\end{equation}
with $D$ the dimension of the space in which the code is embedded.\\

\begin{figure*}
  \centering
  \includegraphics[width=0.8\textwidth]{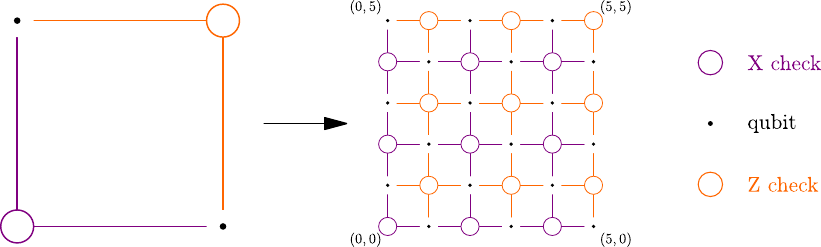}
  \caption{The subdivision of a square of the code.}
  \label{fig:subdivision_balanced_product_code}
\end{figure*}

Recently, it was shown \cite{portnoy2023local, lin2023geometrically, williamson2023layer, li2024transform} how to convert a good quantum LDPC code into a geometrically local code that saturates the BPT bound. Focusing on \cite{lin2023geometrically}, the key idea is to use the local product structure that most of the good quantum LDPC codes possess to represent the code as a square complex (a set of vertices, edges and squares) and apply a subdivision process to locally embed it in $\mathbb{Z}^D$. Finally, to obtain the geometrically local code, the subdivided square complex is again represented as a chain complex. Intuitively, the code is made local by forcing an embedding in $\mathbb{Z}^D$ and removing any long-range connections by subdividing them with additional qubits and checks.\\
The subdivision works as follows. Pick an odd number $L$ and subdivide each square of the square complex into an $L\times L$ grid. Place a Cartesian coordinate system $(i,j)$ on the grid, oriented such that the $X$-check corresponds to $(0,0)$ and the $Z$-check corresponds to $(L,L)$. The subdivided chain complex, $X_L : \mathbb{F}_2^{X_L(0)} \to \mathbb{F}_2^{X_L(1)} \to \mathbb{F}_2^{X_L(2)}$, is then defined by letting
\begin{enumerate}[label=(\roman*)]
    \item $X_L(0)$ (i.e. the $X$-checks) be the set of vertices with $i,j$ both being even.
    \item $X_L(1)$ (i.e. the qubits) be the set of vertices with $i,j$ one being even, one being odd.
    \item $X_L(2)$ (i.e. the $Z$-checks) be the set of vertices with $i,j$ both being odd.
\end{enumerate}
In other words, each square of the code is filled in with a patch of the standard, rectangular surface code. The subdivision of one square of the code is shown in \cref{fig:subdivision_balanced_product_code}.\\

The geometrically local codes from \cite{lin2023geometrically} saturate\footnote{The saturation of the BPT bound was demonstrated up to a polylog factor. However, we are aware of an upcoming result by Lin and Portnoy \cite{lin2024polylog} that removes the polylog factor in the embedding of \cite{portnoy2023local} and consequently in \cite{lin2023geometrically}. We will therefore omit the polylog factors.} the BPT bound [\cref{eq:BPT}] and possess an optimal energy barrier $\mathcal{E} = \Omega\left(n^{\frac{D-2}{D}}\right)$. In terms of the parameters of the underlying good quantum LDPC (qLDPC) code $[[\tilde{n}, \tilde{k}, \tilde{d}]]$, the parameters of the subdivided code are given by

\begin{subequations}
\begin{eqnarray}
n &=& \Theta\left( L^2 \tilde{n}\right)\ ,\label{eq:subdivided_n}\\
k &=& \tilde{k}\ ,\label{eq:subdivided_k}\\
d &=& \Theta\left( L \tilde{d}\right)\ ,\label{eq:subdivided_d}\\
\mathcal{E} &=& \Theta\left(\tilde{\mathcal{E}}\right)\ .
\end{eqnarray}
\end{subequations}


\subsection{Generalized repetition code and generalized surface code}

\begin{figure}[ht]
  \centering
  \includegraphics[width=\linewidth]{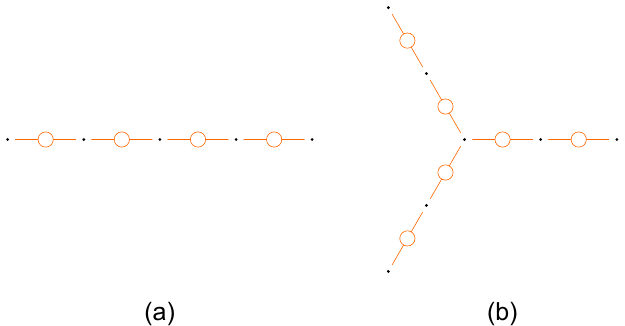}
  \caption{(a) The standard repetition code. (b) The generalized repetition code.}
  \label{fig:gen_rep_code}
\end{figure}
\begin{figure}[hb]
  \centering
  \includegraphics[width=\linewidth]{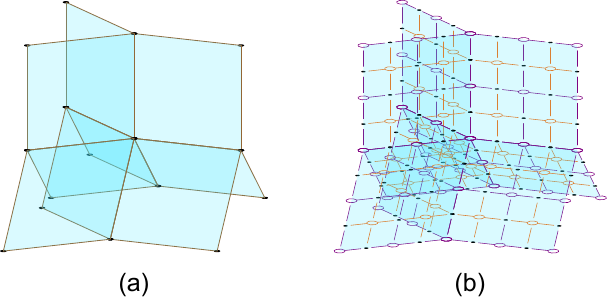}
  \caption{(a) The Cartesian product of two Y-shaped structures with degree 3 branching. (b) The generalized surface code.}
  \label{fig:gen_surf_code}
\end{figure}    

The repetition code consists of a 1D chain of bits where all the bits are required to take the same value. The generalized repetition code is similar, but it may include a branching point in the center \cite{lin2023geometrically}, see \cref{fig:gen_rep_code}. We say that a generalized repetition code has length $L$ if there are $L$ bits on a path from one boundary to another.

Just as the surface code can be defined as the tensor product of two repetition codes, the generalized surface code can be defined as the tensor product of two generalized repetition codes \cite{lin2023geometrically}, as shown in \cref{fig:gen_surf_code}. What we obtain is a collection of squares of the surface code that are joined together along the branches of the two repetition codes. We will refer to these joining branches of the generalized surface code as \emph{seams}, and to the squares of the surface code (without the seams) as \emph{squares}. We say that a generalized surface code has length $L$ if the two generalized repetition codes have length $L$.

Note that the lengths of the generalized repetition code and the generalized surface code are defined in such a way that the length of these codes in the subdivision procedure is given by the subdivision parameter $L$.


\subsection{Structure of the code}

\begin{figure*}
  \centering
  \includegraphics[width=0.8\textwidth]{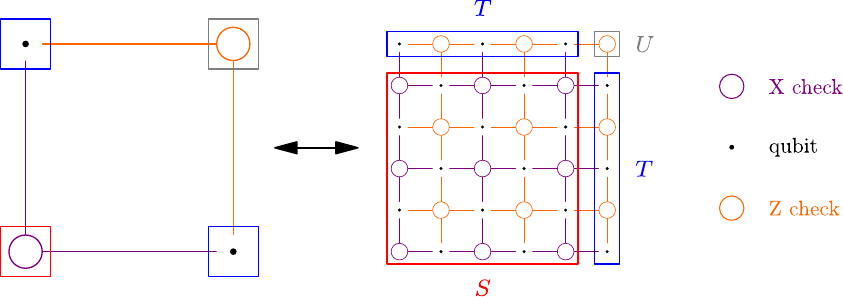}
  \caption{Filling of a square of the balanced product code with an $L\times L$ square patch of the surface code ($L=5$). The regions $S$, $T$ and $U$ in the subdivided square correspond to the $X$-check, qubits and $Z$-check in the original square, respectively.}
  \label{fig:square_division}
\end{figure*} 

The geometrically local code constructed in \cite{lin2023geometrically} is obtained by taking a good qLDPC code which has a balanced product code formulation, and applying a subdivision process to it. In particular, each square in the balanced product code (formed from two qubits and a neighboring $X$- and $Z$-check) is filled in by an $L\times L$ square patch of the surface code, \cref{fig:square_division}.\\

By taking $L$ sufficiently large, namely $L=V^{\frac{1}{D-2}}\log^{D+1}V$, the code becomes geometrically local in dimension $D$ while retaining its optimal parameters. Here $V=\abs{X(0)} + \abs{X(1)} + \abs{X(2)}$ is the total number of code elements, i.e. the sum of the number of $X$-checks, the number of qubits, and the number of $Z$-checks.\\

Because a balanced product code is locally a product, the resulting code is a collection of generalized surface code patches, connected to each other through the $T$-regions, which correspond to the qubits in the original code. A very simple 2D example is shown in \cref{fig:multiple_patches}. In general, the patches are generalized surface codes, and their boundary (the $T$-region) is a generalized repetition code.


\subsubsection{Logical errors}\label{sec:logical errors}

Although the subdivided code can be seen as the combination of a good qLDPC outer code and a surface inner code, this can be misleading since it is not a concatenation and the patches of generalized surface codes do not contain any logical information on their own. This follows from the exactness of the chain complex of the generalized surface code (Corollary 5.5 in Ref. \cite{lin2023geometrically}).

Even though there are no nontrivial logical operations within a single patch, a string\footnote{A string must be interpreted here as a 1D structure that may include branching points or loops.} of errors crossing many patches is able to form a logical operation. It is therefore important to avoid the creation of such error strings during the decoding process.\\

Furthermore, since $d = \Theta(L \tilde{d}) = \Theta(L \tilde{n}) = \Theta\left[L (\text{no. patches})\right]$ [\cref{eq:subdivided_d}], a decoder that works up to a constant fraction of the distance must be able to handle a number of errors per patch, which is linear in $L$.


\section{Decoder of Subdivided Code}\label{sec: Decoder of Subdivided Code}

The strategy to decode the subdivided code is to first push the errors from within each patch to the boundary $T$-regions, and then to push the errors from the $T$-regions to the corner $U$ vertices. In this way, the syndrome is fully supported on the $U$ vertices, which allows the use of the decoder of the underlying good qLDPC code \cite{gu2023efficient,leverrier2023efficient,dinur2023good}. The reason for this is the one-to-one correspondence of the $S$, $T$ and $U$ regions in the subdivided code and the $X$-checks, qubits and $Z$-checks in the qLDPC code, respectively. The correction predicted by the qLDPC decoder is therefore applied to the $T$-regions of the corresponding erroneous qubits.\\

The full decoder of the subdivided code consists of three sub-decoders, each acting on a different region:
\begin{enumerate}
    \item A Union-Find decoder for the generalized surface code, acting on the $S$ regions.
    \item A minimum weight perfect matching decoder for the generalized repetition code, acting on the $T$ regions.
    \item The decoder of the underlying good quantum LDPC code, acting on the $U$ regions.
\end{enumerate}

An overview of the decoder of the subdivided code is given in \cref{alg:decoder_subdivided}.

\begin{algorithm}[H]
    \caption{Decoder for subdivided code}\label{alg:decoder_subdivided}
    \hspace*{\algorithmicindent} \textbf{Input\quad :} The syndrome $\sigma$ of an error $E_X$. \\
    \hspace*{\algorithmicindent} \textbf{Output\quad :} An estimation $\mathcal{C}$ of $E_X$ up to a stabilizer.
    \begin{algorithmic}[1]
        \For {every patch of generalized surface code, $\mathcal{S}_i$}
            \State Apply the Union-Find generalized surface code decoder, as given in \cref{alg:UF-lin}.
        \EndFor
        \For {every boundary $T$-region, $\mathcal{T}_i$}
            \State Apply the minimum weight perfect matching decoder for the generalized repetition code, as given in \cref{alg:MWPM_gen_rep}.
        \EndFor
        \State Use the decoder of the underlying good qLDPC code to the $U$ vertices, and apply the correction to the corresponding $T$-regions.
    \end{algorithmic}
\end{algorithm}

The serial decoder of the subdivided code runs in almost linear time and can correct errors up to a constant fraction of the distance.

\begin{theorem}\label{thm:subdivided decoder}
    The decoder for the subdivided code, as described in \cref{alg:decoder_subdivided}, runs in almost linear time $O(n\alpha(n))$ and can correct an $X$-error of weight $s$ if $s \leq \frac{(L-1)r}{4}$.
\end{theorem}

\begin{remark}
    A simple parallel version of the subdivided decoder (for adversarial noise) is obtained by doing the following:
    \begin{enumerate}[label=(\arabic*)]
    \item Applying the Union-Find decoder to each patch of the generalized surface code in parallel: $O(n^{2/D})$.
    \item Applying the MWPM decoder to each $T$-region in parallel: $O(n^{1/D})$.
    \item Applying the parallel version of qLDPC decoder \cite{leverrier2023decoding,gu2024single}: $O(\log(n))$.
\end{enumerate}
Hence the time-complexity of this parallel decoder is $O(n^{2/D})$. Moreover, since there is no known decoder for the surface code with a sublinear worst-case time-complexity for adversarial noise, this appears to be the most one can achieve.
\end{remark}


\section{Minimum Weight Perfect Matching of Generalized Repetition Code}

In this section, we describe the minimum weight perfect matching decoder that is used on the $T$-regions of the subdivided code in \cref{alg:decoder_subdivided}.

Since the generalized repetition code has a 1D structure, the state of a syndrome bit is fully determined by the error on the qubit to its left and the error on the qubit to its right. Fixing the error on the central qubit therefore fixes the error on all qubits. To obtain the minimum weight perfect matching, we simply select the configuration (error or no error on the central qubit) with the minimum number of errors.

\cref{alg:MWPM_gen_rep} describes this procedure.

\begin{algorithm}[H]
    \caption{Minimum weight perfect matching decoder of the generalized repetition code}\label{alg:MWPM_gen_rep}
    \hspace*{\algorithmicindent} \textbf{Input\quad :} A generalized repetition code with a branching point of degree $\Delta$, and the syndrome $\sigma$ of an error $E_X$. \\
    \hspace*{\algorithmicindent} \textbf{Output\quad :} A minimum weight error $\mathcal{C}$ producing the syndrome $\sigma$.
    \begin{algorithmic}[1]
        \State Place an $X$-error on the central (hyper)edge.
        \For {all arms of the code}
            \For {all edges $i$ ($i$ increases away from the center)}
                \If {edge $i$ is preceded by a syndrome vertex}
                    \State Place the same error on $i$ as on $i-1$
                \Else
                    \State Place the same error on $i$ as on $i-1$ plus an additional $X$-error
                \EndIf
            \EndFor
        \EndFor
        \If {the number of placed errors is $\leq$ half the number of (hyper)edges}
            \State \Return the placed errors.
        \Else
            \State \Return the placed errors with an additional $X$-error on each (hyper)edge.
        \EndIf
    \end{algorithmic}
\end{algorithm}

\begin{theorem}\label{thm: MWPM gen rep}
    The minimum weight perfect matching decoder for the generalized repetition code, as described in \cref{alg:MWPM_gen_rep}, runs in linear time and can correct an $X$-error of weight $s$ if $s \leq \frac{(L-1)\Delta}{4}$, with $\Delta$ the degree of the branching point.
\end{theorem}

\begin{proof}
   It is clear that \cref{alg:MWPM_gen_rep} runs in linear time since we only need to run over the (hyper)edges and vertices once. Since the errors on both sides of a vertex differ if and only if the vertex belongs to the syndrome, the algorithm produces an error with the given syndrome $\sigma$. Because the placement of an error on the central (hyper)edge determines the error on all the other edges, there are only two possible error configurations. Since these configurations differ on each (hyper)edge, their weights add up to the total number of (hyper)edges, which is also the distance of the code: $d = \frac{(L-1)\Delta}{2} + 1$. Finally, because we pick the error configuration with the smallest weight, the decoding works for $X$-errors with a weight of up to $\frac{d-1}{2} = \frac{(L-1)\Delta}{4}$.
\end{proof}


\section{Generalized Erasure Decoder}

\begin{figure*}
  \centering
  \includegraphics[width=0.8\textwidth]{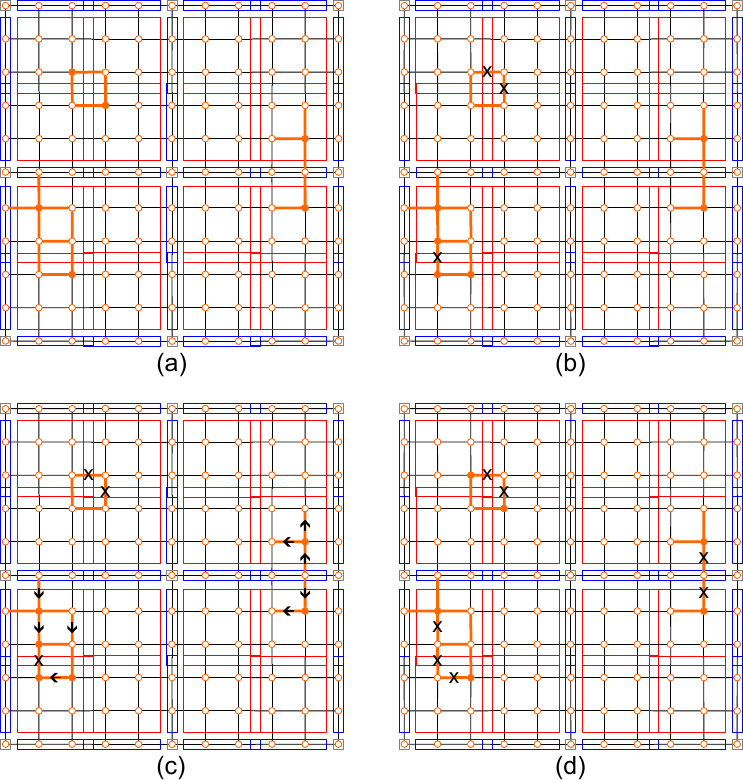}
  \caption{Generalized erasure decoder applied to four patches of the generalized surface code. The algorithm runs from a to d. (a) The erasure (hyper)edges and the syndrome vertices are marked in orange. (b) $X$-errors are placed on the satisfying (hyper)edges and their vertices are flipped in $\sigma$. (c) Spanning trees are constructed in each square. The arrows indicate the growing direction of the tree. (d) The trees are peeled away from their leaves and the $X$-error is constructed. The original syndrome vertices are again indicated in orange to show that the estimated $X$-error produces the syndrome.}
  \label{fig:gen_erasure_decoder}
\end{figure*}

In this section we present an adaptation of the erasure decoder \cite{delfosse2020linear} such that it can be applied to the generalized surface code. This generalized erasure decoder will be needed in \cref{sec: Generalized Union-Find Decoder} to construct a Union-Find decoder for the generalized surface code. As is the case for the regular Union-Find decoder \cite{delfosse2021almost}, the Union-Find decoder for the generalized surface code consists of two parts. First the error is reduced to an erasure error, and second, the erasure error is decoded. In this section we will thus focus on the second part.\\

To apply the erasure decoder to the generalized surface code, few adaptations need to be made. First, the cluster erasure must be split into subclusters, each of which is only contained within a single square of the generalized surface code. In this way, the regular surface code erasure decoder with boundary may be applied to each subcluster. However, this only works if after splitting, each sub-cluster is \textit{satisfiable}, that is, it supports an error that is able to produce the syndrome on the sub-cluster.\\

To achieve satisfiable subclusters, we do the following. First, we determine for each square $s$ that supports some part of the cluster whether or not the restriction of the cluster to that square $C\vert_s$ is connected to the boundary. If it is connected to the boundary, $C\vert_s$ forms a satisfiable subcluster. If it is not connected to the boundary, $C\vert_s$ is satisfiable if and only if the parity of the syndrome on $C\vert_s$ is even.\\

To make the sub-clusters with an odd parity satisfiable, we give them an additional syndrome vertex by placing errors on some of the (hyper)edges on the seams of the generalized surface code. Note that we have to place at most one error on each seam to achieve the change in parity. We must also be careful not to alter the parity of the even sub-clusters (not connected to the boundary) during this process.\\

We are therefore looking for a solution $x$ of the equation
\begin{equation}\label{eqn: satisfiability}
    \sum_{i}x_iG_{is} = p_s
\end{equation}
for all squares $s$ that are not connected to the boundary. The components of \cref{eqn: satisfiability} are defined as follows:
\begin{enumerate}[label=(\roman*)]
    \item $x$ is a row vector in $\mathbb{F}_2$ with $x_i = 1$ ($x_i = 0$) indicating that seam $i$ has an error (no error).
    \item $G$ is a matrix in $\mathbb{F}_2$ with $G_{is} = 1$ if and only if seam $i$ neighbors square $s$.
    \item $p$ is a row vector in $\mathbb{F}_2$ with $p_s = 1$ if and only if the parity of the syndrome on $C\vert_s$ is odd.
\end{enumerate}

Because the errors on the (hyper)edges on the seams need to be contained within the cluster erasure, we should only look for a solution $x$ that is supported on the seams with some overlap with the cluster erasure. Because the original cluster erasure is by definition satisfiable, there will always be a solution $x$ of \cref{eqn: satisfiability} that is contained within the cluster erasure. We refer to a set of (hyper)edges that implement $x$ as a \textit{satisfying configuration}.\\

Splitting the cluster erasure into subclusters is then done by removing all the (hyper)edges on the seams, and flipping the vertices of the (hyper)edges of the satisfying configuration in $\sigma$. Finally, the error of the cluster erasure is obtained by taking the union of the errors on the (hyper)edges of the satisfying configuration and the errors we get by applying the regular erasure decoder to each square.\\

An overview of the full erasure decoder for the generalized surface code is presented in \cref{alg:erasure} (see \cref{sec:erasure_surf} for the relevant definitions) and \cref{fig:gen_erasure_decoder} illustrates the different steps of the algorithm. Finally, \cref{thm:gen erasure decoder} shows that the algorithm decodes correctly and that it runs in linear time.

\begin{algorithm}
    \caption{Erasure decoder for generalized surface code}\label{alg:erasure}
    \hspace*{\algorithmicindent} \textbf{Input\quad :} A generalized surface code with boundary, a cluster erasure $\varepsilon \subset \mathring{E}$, and the syndrome $\sigma \subset \mathring{V}$ of an $X$ error. \\
    \hspace*{\algorithmicindent} \textbf{Output\quad :} An $X$-error $P$ such that $P\subset \varepsilon$ and $\sigma(P) = \sigma$.
    \begin{algorithmic}[1]
        \State Initialize $A$ by $A = \emptyset$.
        \State Find a satisfying configuration $x$ of the cluster.
        \State Add the satisfying hyperedges to $A$ and flip their vertices in $\sigma$.
        \For {each square $s$ of the cluster}
            \If {$C\vert_s$ is connected to the boundary}
                \State Grow a spanning tree $F_{\varepsilon\vert_s}$ in $s$ from a vertex on the boundary.
            \EndIf
            \If {$C\vert_s$ is not connected to the boundary}
                \State Grow a spanning tree $F_{\varepsilon\vert_s}$ from any vertex.
            \EndIf
            \While {$F_{\varepsilon\vert_s} \neq \emptyset$}
                \State pick a leaf edge $e=\{ u, v\}$ with pendant vertex $u \in \mathring{V}$, remove $e$ from $F_{\varepsilon\vert_s}$
                \If {$u \in \sigma$}
                    \State add $e$ to $A$, remove $u$ from $\sigma$, and flip $v$ in $\sigma$.
                \EndIf
                \If {$u \not\in \sigma$}
                    \State do nothing.
                \EndIf
            \EndWhile
            \EndFor
        \State Return $P = \prod_{e \in A} Z_e$. 
    \end{algorithmic}
\end{algorithm}

\begin{theorem}\label{thm:gen erasure decoder}
The erasure decoder for the generalized surface code, as described in \cref{alg:erasure}, provides for a given cluster erasure $\varepsilon$ and syndrome $\sigma$ an $X$-error $P$ such that $P \subset \varepsilon$ and $\sigma(P) = \sigma$. Moreover, the decoder runs in linear time.
\end{theorem}

\begin{proof}
We show that the set $A$ satisfies the following properties: $A \subset \varepsilon$ and $\partial (A) = \sigma$, where $\partial (A)$ denotes the set of vertices that $A$ encounters an odd number of times. The statement $A \subset \varepsilon$ is immediate since we only add elements of $\varepsilon$ to $A$.

To show that $\partial (A) = \sigma$, first note that removing all hyperedges, except those in $x$, does not change the satisfiability of the cluster. Furthermore, removing the hyperedges in $x$ while flipping their vertices in $\sigma$, also does not change the satisfiability of the cluster. Call $\sigma'$ the modified syndrome, i.e. $\sigma'(v) = \sigma(v)$ for every vertex $v$ not neighboring an edge in $x$, and $\sigma'(v) = \left[\sigma(v) + j\right]\mod_2$ for every vertex $v$ neighboring $j$ hyperedges in $x$.

Since each connected component of the cluster is now fully contained within a square $s$, the satisfiability of the cluster implies that for each square $s$, $C\vert_s$ is satisfiable. The regular erasure decoder \cite{delfosse2020linear} can therefore be applied to each subcluster $C\vert_s$ with syndrome $\sigma'\vert_s$, which produces $A'\vert_s$ with $A'\vert_s \subset \varepsilon\vert_s$ and $\partial (A'\vert_s) = \sigma'\vert_s$.

Because $A = \bigcup_{e \in x}\{e\} \cup \bigcup_s A'\vert_s$, its boundary evaluated in a vertex $v$ is given by
\begin{equation}
    \partial (A)(v) = \left[ \partial\left(\bigcup_{e \in x}\{e\}\right) (v) + \sum_s \partial\left(A'\vert_s\right) (v)\right]\mod_2\ .
\end{equation}
If $v \in \tilde{s}$ does not neighbor any hyperedge in $x$, we find $\partial (A)(v) = \partial\left( A'\vert_{\tilde{s}} \right) (v) = \sigma'\vert_{\tilde{s}} (v) = \sigma (v)$. If $v \in \tilde{s}$ neighbors some hyperedges $\tilde{e}_1, \ldots, \tilde{e}_j$, we find $\partial (A)(v) = \left[ \partial \left( \bigcup_{e\in\{\tilde{e}_1, \ldots, \tilde{e}_j\}} \{e\}\right) (v) + \partial\left( A'\vert_{\tilde{s}} \right) (v)\right] \mod_2 = \left[ j + \sigma'\vert_{\tilde{s}} (v)\right] = \left[j + \sigma(v) + j\right]\mod_2 = \sigma(v)$. $A$ therefore satisfies the required properties.

We now show that the algorithm runs in linear time. To find a configuration $x$ that satisfies the cluster, we first check every vertex and (hyper)edge to find the squares included in the cluster, the parity of each square, and whether or not a square is connected to the boundary. Finding $x$ from this can then be done in constant time since the number of squares is bounded. Steps (1--3) can therefore be performed in linear time. The rest of the algorithm consists of applying the regular erasure decoder --- which runs in linear time --- to a bounded number of squares. The overall complexity is therefore linear.
\end{proof}


\section{Generalized Union-Find Decoder}\label{sec: Generalized Union-Find Decoder}

In this section, we present a generalized version of the Union-Find decoder \cite{delfosse2021almost} that can be applied to the generalized surface code. In the overall decoder of the subdivided code (\cref{sec: Decoder of Subdivided Code}), this generalized Union-Find decoder is applied to every patch of generalized surface code.\\

To apply the Union-Find decoder to the generalized surface code, some modifications need to be made. As with the generalized erasure decoder, we must upgrade the notion of an odd cluster to an unsatisfiable cluster, i.e. a cluster for which \cref{eqn: satisfiability} does not have a solution. The growing and fusing of edges must also be generalized to include hyperedges. Because we grow a (hyper)edge from a particular vertex towards all the other vertices of the (hyper)edge, growing twice by half a (hyper)edge adds the full (hyper)edge. Similarly, when two half-grown (hyper)edges meet, the full (hyper)edge gets added.\\

\cref{alg:UF-naive} gives an overview of the Union-Find decoder for the generalized surface code. Note that the output of the algorithm is an estimation of the error up to a stabilizer, but excluding stabilizers that turn a non-patch-crossing path into a patch-crossing path. This difference with the Union-Find decoder for the surface code is due to the fact that the generalized surface code on the $S$ region does not contain any logical information (\cref{sec:logical errors}). Although the logical state of the generalized surface cannot change by crossing the patch, it can affect the logical state of the subdivided code. The output of \cref{alg:UF-naive} is therefore exactly what is needed.\\

\cref{thm:UF_gen} is the analog of \cref{thm:UF decoder surface}; it shows that the generalized Union-Find decoder can correct $t$ erased qubits and $s$ Pauli errors if $t + 2s < L$.\\

\begin{algorithm}
    \caption{Union-Find generalized surface code decoder - Naive version}\label{alg:UF-naive}
    \hspace*{\algorithmicindent} \textbf{Input\quad :} The set of erased positions $\varepsilon \subset E$ and the syndrome $\sigma$ of an error $E_X$. \\
    \hspace*{\algorithmicindent} \textbf{Output\quad :} An estimation $\mathcal{C}$ of $E_X$ up to a stabilizer.\textsuperscript{\hyperlink{fn:a}{\textcolor{blue}{a}}}
    \begin{algorithmic}[1]
        \State Create the list of all unsatisfiable clusters $C_1, \ldots, C_m$, and initialize the modified erasure $\varepsilon' = \varepsilon$.
        \While {there exists an unsatisfiable cluster}
            \For {all unsatisfiable clusters $C_i$}
                \State Grow $C_i$ by increasing its radius by one half-(hyper)edge.
                \If {$C_i$ meets another cluster}
                    \State fuse and update satisfiability.
                \EndIf
                \If {$C_i$ meets the boundary}
                    \State update satisfiability.
                \EndIf
                \If {$C_i$ is satisfiable}
                    \State remove it from the unsatisfiable cluster list.
                \EndIf
            \EndFor
        \EndWhile
        \State Add full (hyper)edges that are in the grown clusters to $\varepsilon'$.
        \For {all satisfiable clusters}
            \State apply the erasure decoder for the generalized surface code to find $\mathcal{C}$.
        \EndFor
    \end{algorithmic}
    \vspace{0.5em}
    \hrule
    \vspace{0.2em}
    \hrule
    \vspace{0.3em}
    {\raggedright\footnotesize\hypertarget{fn:a}{} \textsuperscript{a}Excluding stabilizers that turn a non-patch-crossing path into a patch-crossing path.\par}
\end{algorithm}

\begin{theorem}\label{thm:UF_gen}
The Union-Find decoder for the generalized surface code, as described in \cref{alg:UF-naive}, can correct any combination of t erased qubits and s $X$-errors if $t + 2s < L$.
\end{theorem}

\begin{proof}
With every growing step, an additional half-(hyper)edge of the error is covered by the clusters. This means that there can be at most $2s$ growing steps, leading to a combined diameter growth of at most $2s$ (hyper)edges. In case a boundary is hit, the growth of the clusters can stop early, say after $\tilde{s}$ steps, without the clusters covering the error $E_X$. The largest possible diameter\footnote{The diameter of an error is the minimum sum of diameters of clusters needed to cover the error.} of an error string ($E_X$ and estimated correction) is then $2\tilde{s} + (s-\tilde{s}) = s + \tilde{s} \leq 2s$. Adding the initially $t$ erased qubits bounds the diameter of any error string by $2s + t < L$. This is not enough to cross the patch.
\end{proof}

We now show that the almost-linear time implementation of the Union-Find decoder can also be adapted to the generalized surface code without increasing its time-complexity.

\begin{algorithm}
    \caption{Union-Find generalized surface code decoder - Almost linear time version}\label{alg:UF-lin}
    \hspace*{\algorithmicindent} \textbf{Input\quad :} The set of erased positions $\varepsilon \subset E$ and the syndrome $\sigma$ of an error $E_X$. \\
    \hspace*{\algorithmicindent} \textbf{Output\quad :} An estimation $\mathcal{C}$ of $E_X$  up to a stabilizer.\textsuperscript{\hyperlink{fn:b}{\textcolor{blue}{b}}}
    \begin{algorithmic}[1]
        \State Initialize cluster-trees, Support and boundary lists for all clusters.
        \State Create the list $\mathcal{L}$ of roots of unsatisfiable clusters.
        \While {$\mathcal{L} \neq \emptyset$}
            \State Initialize fusion list $\mathcal{F} = \emptyset$
            \For {all $u \in \mathcal{L}$}
                \State Grow the cluster $C_u$ a half (hyper)edge in the Table Support.
                \If {a new grown boundary edge $e$ is added in Support}
                    \State Update boundary at root $u$.
                \EndIf
                \If {a new grown interior (hyper)edge $e$ is added in Support}
                    \State Add $e$ to the fusion list $\mathcal{F}$.
                \EndIf
            \EndFor
            \For {all $e = \{v_1, \ldots, v_l\} \in \mathcal{F}$}
                \State Determine Find($v_i$).
                \If {there are at least two distinct roots}
                    \State Read the sizes at the distinct roots.
                    \State Union the smaller clusters to the biggest cluster.
                    \State Append the boundary lists of the smaller clusters at the end of the boundary list of the biggest cluster.
                \EndIf
                \If {all roots are identical}
                    \State Remove $e$ from the list $\mathcal{F}$.
                \EndIf
            \EndFor
            \For {each $u \in \mathcal{L}$}
                \If {$Find(u) \neq u$}
                    \State remove $u$ from $\mathcal{L}$.
                \EndIf
            \EndFor
            \For {all $u \in \mathcal{L}$}
                \State Remove the vertices of the boundary lists of $u$ that are not boundary vertices.
            \EndFor
            \For {all $u \in \mathcal{L}$}
                \State Calculate the satisfiability of $C_u$.
                \If {$C_u$ is satisfiable}
                    \State Remove $u$ from $\mathcal{L}$.
                \EndIf
            \EndFor
        \EndWhile
        \State Erase all the (hyper)edges that are fully grown in Support.
        \State Apply the erasure decoder for the generalized surface code.
    \end{algorithmic}
    \vspace{0.5em}
    \hrule
    \vspace{0.2em}
    \hrule
    \vspace{0.3em}
    {\raggedright\footnotesize\hypertarget{fn:b}{} \textsuperscript{b}Excluding stabilizers that turn a non-patch-crossing path into a patch-crossing path.\par}
\end{algorithm}

\subsection{Data structures}

The data structures in \cref{alg:UF-lin} are largely the same as those in \cite{delfosse2021almost}. However, few modifications are needed to keep track of the additional structure of the clusters in the generalized surface code.

\paragraph{Cluster-tree}

A cluster-tree is a tree whose vertices correspond to the vertices of a cluster $C$. It is initialized by taking an arbitrary vertex $u$ of $C$ as the root of the tree, and adding all other vertices of $C$ at depth $1$.\\

\noindent The root stores for each square $s$ 
\begin{itemize}
    \item the \textit{square inclusion}: does $s$ contain vertices of $C$
    \item the \textit{parity} of $\sigma\vert_{s \cap C}$
    \item the \textit{boundary} status: does $C$ connect to the boundary in $s$.
\end{itemize}
The root also stores for each seam $f$
\begin{itemize}
    \item the \textit{seam inclusion}: if $f$ contains hyperedge(s) of $C$, the seam inclusion is any such hyperedge, otherwise the seam inclusion of $f$ is the empty set.
\end{itemize}

\paragraph{Support}

Support is a look-up table that stores for each (hyper)edge its \textit{filled status} and \textit{boundary status}. The filled status can take one of three values: \textit{Unoccupied}, \textit{Half-grown} or \textit{Grown}. While the filled status is updated during the algorithm, the boundary status of a (hyper)edge is fixed and simply indicates whether it connects to the boundary of the patch of the generalized surface code (a \textit{boundary edge}) or not [an \textit{interior (hyper)edge}].

Support is initialized by marking all the erased (hyper)edges as \textit{Grown} and marking all others as \textit{Unoccupied}. Each (hyper)edge is also given its appropriate boundary status.

\paragraph{Boundary List}

The Boundary List of a cluster $C$ is a list of all the boundary vertices of $C$, where a \textit{boundary vertex} of $C$ is a vertex of $C$ with at least one of its incident (hyper)edges not in the erasure $\varepsilon$.\\

\begin{theorem}
    The implementation of the generalized Union-Find decoder, as described in \cref{alg:UF-lin}, runs in almost linear time.
\end{theorem}

\begin{proof}
 \cref{alg:UF-lin} can be divided into three blocks. The first block consists of lines 1--2 and initializes the clusters. The second block consists of lines 3--28 and performs the growing and fusion of clusters. Finally, the third block consists of line 29 and applies the erasure decoder for the generalized surface code.\\ 

 The creation of the cluster-trees can be achieved in linear time by applying the connected components algorithm from \cite{hopcroft1973algorithm} to find the connected components of the sublattice induced by the erased (hyper)edges. This algorithm can directly be applied if we replace an $l$-hyperedge, discovered by the algorithm from vertex $u$, by $l$ edges emanating from $u$, i.e. $\{u, v_1, \ldots, v_l\} \longrightarrow \{u, v_1\}, \ldots, \{u, v_l\}$. During this exploration of the connected components of the erasure, we can also calculate the square inclusion, the parity, and boundary status of the cluster at a constant overhead.
 
 The remaining clusters are those that consist of a single syndrome vertex. The creation of the cluster-tree for each of these types of clusters is achieved in constant time. The overall time-complexity of the initialization of the cluster-trees is thus linear.
 
Support and the Boundary Lists can be initialized in linear time since we only have to check each (hyper)edge and vertex once, and each check takes a constant time.

The list $\mathcal{L}$ of unsatisfiable clusters is created by calculating for each cluster its satisfiablility. Since the number of squares in the generalized surface code is bounded, the satisfiability of a cluster can be calculated from its square inclusion, parity and boundary status in a constant time. Since the number of clusters is in the worst case linear in the number of qubits, the time-complexity of the creation of $\mathcal{L}$ is also linear.

Block 1 of the algorithm can therefore be performed in linear time.\\

Growing the clusters (lines 6--10) is done by running over the vertices in the boundary lists, and for each vertex growing its bounded number of adjacent (hyper)edges. Because each vertex can be a boundary vertex for at most two rounds of growth, this takes at most linear time.

Lines 11--18 mainly describe the Union-Find procedure, which is known to have a time complexity of $O(n\alpha(n))$. The only difference is that we are now also appending the boundary lists of the smaller clusters at the end of the boundary list of the biggest cluster. However, since each vertex can be a boundary vertex for at most two rounds of growth, this does not change the complexity.

The linearity of lines 19--27 is again due to the fact that a vertex is a boundary vertex for at most two rounds of growth, and that a root $u\in \mathcal{L}$ has a nonempty boundary list after each round of growth. This may need some more explanation: The only way a root $u \in \mathcal{L}$ can have an empty boundary list is if $C_u$ covers the full patch of the generalized surface code, such that no further growth is possible. However, such a cluster will be satisfiable and $u$ will therefore be removed from $\mathcal{L}$. After each round of growth, each root $u \in \mathcal{L}$ therefore has a nonempty boundary list. This also means that (after each round of growth) the number of roots is smaller than, or equal to, the number of boundary vertices.

Because a vertex can be a boundary vertex for at most two rounds of growth, the number of roots $u$ that have to be checked in the full algorithm is smaller than three times the number of vertices. (The factor of 3 accounts for the case in which a cluster loses its last boundary vertex during growth.)

Evidently, adding the fully grown (hyper)edges to the erasure (line 28) is done in linear time, which shows that block 2 of the algorithm can be performed in almost linear time $O(n\alpha(n))$.\\

By \cref{thm:gen erasure decoder}, the erasure decoder for the generalized surface code runs in linear time.\\

The complete algorithm has therefore an almost linear time-complexity $O(n\alpha(n))$.
\end{proof}


\section{Proof of the main theorem}\label{sec:proof of main theorem}

\begin{proof}[Proof of \cref{thm:subdivided decoder}]
    Since the algorithm consists of applying three consecutive decoders, which all run in almost linear time or linear time, the overall time-complexity is also almost linear.
    
    We now show that the decoder works for error weights up to $\frac{(L-1)r}{4}$, where $r$ is the maximum number of errors that the decoder of the underlying good qLDPC code is guaranteed to correct.\\

    There are two main reasons why the decoder may fail:
    \begin{enumerate}[label=(\arabic*)]
        \item The minimum weight perfect matching decoder for the generalized repetition code fails.
        \item The Union-Find generalized surface code decoder fails.
    \end{enumerate}
    
    We first consider the failure of the minimum weight perfect matching decoder for the generalized repetition code.\\
    
    Note that the largest diameter of the total error (initial error plus decoding error) that can be achieved by applying the Union-Find generalized surface code decoder to an initial error of weight $s$ is at most $2s$ (see the proof of \cref{thm:UF_gen}). Let $B$ be a set of clusters (with minimal total diameter) covering the total error. Then, assuming that $s<\frac{L}{2}$, no cluster in $B$ can neighbor more than two $T$-regions. In this case, the diameter of a cluster forms an upper-bound for the number of errors that can be pushed from that cluster to the $T$-regions.
    
    Therefore, to get more than $\frac{L-1}{2}$ errors in a $T$-region, more than $\frac{L-1}{4}$ errors are needed from the patches. Of course, the $T$-region itself may contain initial errors as well. However, since initial errors in the $T$-region count as singular, and those in the patches count as double, the worst-case scenario is where all the initial errors are in the patches.\\

    Now, if the total number of initial errors in the subdivided code is at most $\frac{(L-1)r}{4}$, there are fewer than $r$ $T$-regions with more than $\frac{L-1}{2}$ errors, and they may not be decoded correctly. This leads to fewer than $r$ errors on the good qLDPC decoding level, which is therefore guaranteed to succeed.\\

    Finally, we consider the failure of the Union-Find generalized surface code decoder.\\

    By \cref{thm:UF_gen}, the Union-Find generalized surface code decoder can only fail when a patch contains more than $\frac{L-1}{2}$ initial errors. However, if the total number of initial errors in the subdivided code is at most $\frac{(L-1)r}{4}$, there are fewer than $\frac{r}{2}$ patches with more than $\frac{L-1}{2}$ initial errors, and they may not be decoded correctly. This leads to fewer than $\frac{r}{2}$ errors on the good qLDPC decoding level, which is therefore guaranteed to succeed.
\end{proof}


\section{Threshold}

In this section we consider the effect of random noise. More specifically, we look at the code under the code capacity noise model.

\subsection{Threshold of the good qLDPC code}

Given a good qLDPC code with a decoder that decodes correctly up to $r$ errors, the probability of a logical error is given by
\begin{subequations}
    \begin{eqnarray}
        P_L &\leqslant& P(\text{no. errors} \geqslant r + 1)\\
        &=& \sum_{i=r+1}^{n_{\text{qLDPC}}} \binom{n_{\text{qLDPC}}}{i}p^i (1-p)^{n_{\text{qLDPC}}-i}\\
        &=& F(n_{\text{qLDPC}}-(r+1);n_{\text{qLDPC}},1-p)
    \end{eqnarray}
\end{subequations}
with $F(r+1;n_{\text{qLDPC}},p)$ the cumulative distribution function of the binomial distribution. By taking $n_{\text{qLDPC}}p \leqslant r+1$, Hoeffding's inequality yields the bound
\begin{equation}
    P_L \leqslant \exp\left[-2n_{\text{qLDPC}}\left(\frac{r+1}{n_{\text{qLDPC}}} - p\right)^2\right].
\end{equation}
This upper bound can be made arbitrarily small by increasing $d\propto n_{\text{qLDPC}}$, provided $p \leqslant \frac{r+1}{n_{\text{qLDPC}}}$.

This shows that a good qLDPC code always has a threshold under the code capacity noise model.

\subsection{Threshold of the generalized surface code and generalized repetition code (MWPM $+$ MWPM)}\label{sec:threshold(MWPM+MWPM)}

\begin{figure}
  \centering
  \includegraphics[width=\linewidth]{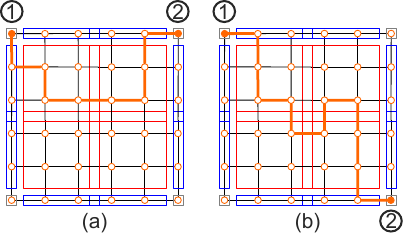}
  \caption{``Logical'' errors of the combined generalized surface code and generalized repetition code. (a) A path connecting the $U$-vertices at the boundary of a single (top) $T$-region. (b) A path connecting the $U$-vertices at the boundary of two (top and right, or, bottom and left) $T$-regions.}
  \label{fig:logical_box}
\end{figure}

We now look at the ``threshold'' of the combined generalized surface code and generalized repetition code, using minimum weight perfect matching on both the generalized surface code and the generalized repetition code. Since these codes do not contain any logical information themselves, it is technically impossible to have a logical error. However, since the $T$-regions correspond to the qubits in the good qLDPC code (see \cref{fig:square_division}), we can identify the logical errors with paths that connect the $U$-vertices at the boundary of one or multiple $T$-regions. Two examples of such paths are shown in \cref{fig:logical_box}.\\

Consider such a path between $U$-vertices. This path can consist of both initial errors and decoding errors, and it can have sections in $T$-regions as well as sections in $S$-regions. Observe that while a segment of the path within a $T$-region can be entirely composed of decoding errors, a segment within an $S$-region can contain decoding errors in at most half of its (hyper)edges (MWPM in $S$). The ratio of decoding errors in the path is therefore maximized by maximizing the number of (hyper)edges of the path within $T$-regions.\\
Since MWPM is also performed on the $T$-regions, each $T$-region can contain decoding errors in at most half of its (hyper)edges. To connect to the next $T$-region (or simply to the $U$-vertices at the endpoints of the path), the remaining errors must be allocated to the $S$-regions. Consequently, the length of the path within $S$ must be at least equal to the length within $T$. Therefore, the maximum proportion of decoding errors in the entire path is $\leq \frac{1}{2} + \frac{1}{4} = \frac{3}{4}$.\\
Focusing on a logical error associated with a single $T$-region, there must be at least one path of $m \geq \frac{(L-1)\Delta}{2} + 1$ (hyper)edges connecting the corner $U$-vertices adjacent to this $T$-region, and containing at least $\ceil{m/4}$ (hyper)edges associated with errors.\\
Since the endpoints of the path are fixed, we have to make at most $m$ choices for the next (hyper)edges. Because each vertex is connected to at most $\Delta$ (hyper)edges and because we don't allow backtracking, the total number of paths of length $m$ (with fixed endpoints) is upper-bounded by $\left(\Delta - 1\right)^m$.\\
Following \cite{fowler2012proof}, we find that for a path of length $m$, the probability of at least $\ceil{m/4}$ of its (hyper)edges being associated with errors is at most $2^m \epsilon^{\ceil{m/4}}$, with $\epsilon$ being the error probability at a single (hyper)edge.\\

The probability of having a ``logical error'' associated with a single $T$-region is therefore
\begin{subequations}
    \begin{eqnarray}
        P_L &\leq& \sum_{m=\frac{(L-1)\Delta}{2} + 1}^{\infty} a^{m} 2^m \epsilon^{\ceil{m/4}}\\
        &\leq& \sum_{m=\frac{(L-1)\Delta}{2} + 1}^{\infty} \left(2a \sqrt[4]{\epsilon}\right)^m\\
        &\leq& \left(2a\sqrt[4]{\epsilon}\right)^{\frac{(L-1)\Delta}{2} + 1} \sum_{m=0}^\infty \left(2a\sqrt[4]{\epsilon}\right)^m\\
        &=& \left(2a\sqrt[4]{\epsilon}\right)^{\frac{(L-1)\Delta}{2} + 1} \frac{1}{1 - 2a\sqrt[4]{\epsilon}}\ .
    \end{eqnarray}
\end{subequations}
Provided that $\epsilon < 1/(2a)^4$, the probability can be made arbitrarily small by increasing $n$ and, consequently, $L$. This establishes the existence of a threshold for the generalized surface and repetition code.
When combined with the threshold of the outer good qLDPC code, this further demonstrates the existence of a threshold for the total subdivided code.\\

\noindent We have therefore proven the following.

\begin{theorem}
    Geometrically local codes from subdivision, together with a combined minimum weight perfect matching and good qLDPC decoder, have a finite threshold error rate. Moreover, logical errors are suppressed exponentially with the subdivision parameter $L$.
\end{theorem}

\subsection{Threshold of the generalized surface code and generalized repetition code (generalized UF $+$ MWPM)}

In this section we discuss the ``threshold'' of the combined generalized surface code and generalized repetition code, using the  generalized Union-Find decoder and the minimum weight perfect matching decoder, respectively.\\

While a formal proof of the existence of a threshold is beyond the scope of this work, we offer an intuitive argument to suggest why this might hold true, noting that a similar level of rigor is also absent for the union-find decoder of the regular surface code.\\

One argument that can be made is the following. Looking back at the proof presented in \cref{sec:threshold(MWPM+MWPM)}, we see that the minimum weight perfect matching on the $T$-regions essentially halves the length required in the $S$-regions to create a logical error. Alternatively, we can interpret this by maintaining the same path length in the $S$-regions (treating the contribution of the $T$-region as part of $S$) and rescaling the number of initial errors needed to construct the path. Since the $T$-region can take up to half the (hyper)edges of the path, the probability of a particular path increases by up to the square root of the original probability, i.e. $\mathbb{P}(\text{path}) = p \rightarrow \mathbb{P}(\text{path}) = \sqrt{p}$. 

Given that numerical evidence suggests that the surface code exhibits similar thresholds for MWPM and the Union-Find decoder \cite{delfosse2021almost}, it is reasonable to infer that this square root reduction in path probability also applies to the Union-Find decoder.

Finally, because the generalized Union-Find decoder can be understood as applying the Union-Find decoder to each square $\squares$, the numerical threshold of the Union-Find decoder for the surface code \cite{delfosse2021almost}, combined with the square root probability reduction (which correspondingly rescales the threshold by the square root), leads to a threshold for the generalized Union-Find and MWPM decoders.

\section*{Acknowledgment}

Q.E. acknowledges the support of the Research Foundation-Flanders through the Fundamental Research PhD programme (grant no. 11Q4A24N), as well as the EOS-FWO-FNRS project CHEQS.

\appendix

\section{Erasure Decoder for the Surface Code}\label{sec:erasure_surf}

\begin{figure*}
  \centering
  \includegraphics[width=0.8\textwidth]{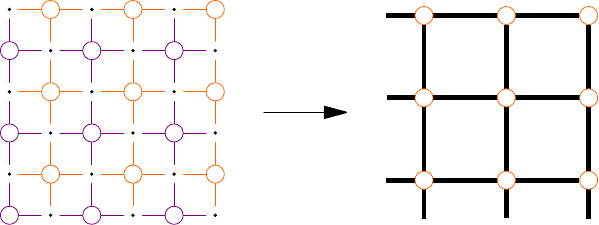}
  \caption{Gridlike representation.}
  \label{fig:grid_representation}
\end{figure*} 

The goal of the erasure decoder for the surface code is to find a $Z$-error and an $X$-error supported on the erasure $\varepsilon$ that reproduce the given syndromes $\sigma_z$ and $\sigma_x$. As stated in \cref{sec:quantum_erasures}, this corresponds to the most likely correction. From now on, and for the rest of the paper, we will focus only on the $X$-errors since the $Z$-errors can be handled in a completely analogous way. The syndrome $\sigma$ should therefore always be understood as the $X$-part of the syndrome, $\sigma_x$.\\

To describe the algorithm, it is best to use a gridlike representation of the surface code whereby the vertices $V$ correspond to the $Z$ checks, the faces correspond to the $X$ checks, the edges $E$ correspond to qubits, and two vertices are connected by an edge if there is a qubit that is acted on by the corresponding checks.  An example of such a gridlike representation is shown in \cref{fig:grid_representation}.\\

The strategy \cite{delfosse2020linear} is to translate the decoding problem (how to pair up the syndrome vertices through paths in the erasure) into a graphical language. To this end, the \textit{boundary} of a set of edges $A$, denoted as $\partial A$, is defined as the set of vertices that $A$ encounters an odd number of times. Decoding is therefore equivalent to finding a set $A$ such that $A \subset \varepsilon$ and $\partial (A) = \sigma$.\\

Finding such a set $A$ in linear-time is made difficult by the many cycles in $\varepsilon$. To avoid this issue, a \textit{spanning forest} $F_\varepsilon$ is created inside $\varepsilon$, which is defined as the maximal subset of edges of $\varepsilon$ that contains no cycles and spans all the vertices of $\varepsilon$. If the erasure $\varepsilon$ together with its neighboring vertices forms a connected graph, $F_\varepsilon$ is also connected and is called a \textit{spanning tree}. A spanning forest can always be found in linear time \cite{cormen2009introduction}.\\

The next step is to peel away the forest $F_\varepsilon$ from its leaves and construct the set $A$ in the process. A \textit{leaf} is defined as an edge $e=\{u, v\}$ that is only connected to the forest by one of its vertices, say $v$. The other vertex, $u$, is called the \textit{pendant vertex}.\\
Starting from the empty set, $A$ is built up by recursively applying the following rules:
\begin{enumerate}[label=(\roman*)]
    \item Pick a leaf edge $e = \{u, v\}$ with pendant vertex $u$ and remove $e$ from $F_\varepsilon$.
    \item If $u \in \sigma$, then add $e$ to $A$, remove $u$ from $\sigma$, and flip $v$ in $\sigma$.
    \item If $u \not\in \sigma$, then do nothing.
\end{enumerate}
Flipping a vertex $v$ in $\sigma$ means that if $v \in \sigma$, $v$ gets removed from $\sigma$, and if $v \not\in \sigma$, $v$ gets added to $\sigma$.\\

Once the forest $F_\varepsilon$ is fully peeled, $A$ is complete and satisfies the requirements $A \subset \varepsilon$ and $\partial (A) = \sigma$. The reason for this is the following: since $F_\varepsilon$ spans all the vertices of $\varepsilon$ and $F_\varepsilon$ gets fully peeled, each vertex is a pendant vertex at some step in the peeling process. If a vertex $u \in \sigma$, that vertex will have a single incident edge of $A$ (rule 2) such that $u \in \partial A$. Alternatively, if $u \not\in \sigma$, $u$ will have no incident edges of $A$ (rule 3) such that $u \not\in \partial A$. For a full proof we refer to \cite{delfosse2020linear}.\\

\cref{alg:Erasure_decoder_surf} summarizes the decoding procedure for a general surface code $G = (V, E, F)$ without boundary.

\begin{algorithm}[H]
    \caption{Erasure decoder for the surface code}\label{alg:Erasure_decoder_surf}
    \hspace*{\algorithmicindent} \textbf{Input\quad :} A surface $G=(V,E,F)$, an erasure $\varepsilon \subset E$ and the syndrome $\sigma \subset V$ of an $X$-error. \\
    \hspace*{\algorithmicindent} \textbf{Output\quad :} An $X$-error $P$ such that $P \subset \varepsilon$ and $\sigma(P) = \sigma$.
    \begin{algorithmic}[1]
        \State Initialize $A$ by $A = \emptyset$.
        \State Construct a spanning forest $F_\varepsilon$ of $\varepsilon$.
        \While{$F_\varepsilon \neq \emptyset$}
            \State Pick a leaf edge $e=\{ u,v\}$ with pendant vertex $u$ and remove $e$ from $F_\varepsilon$.
            \If{$u \in \sigma$}
                \State Add $e$ to $A$, remove $u$ from $\sigma$, and flip $v$ in $\sigma$.
            \EndIf
            \If{$u \not\in \sigma$}
                \State Do nothing.
            \EndIf
        \EndWhile
        \State Return $P = \prod_{e \in A} X_e$.
    \end{algorithmic}
\end{algorithm}

\subsection{Surfaces with boundaries}

To deal with surface codes with boundaries, that is, surface codes whose grid representations have edges belonging to a unique face, some modifications need to be made to the algorithm. The reason for this is that the vertices on the boundary do not contain enough information to reconstruct the error starting from those vertices. It is therefore important to make the distinction between the edges and vertices at the boundary and those not at the boundary.\\

To be precise, let $\breve{E}$ be the set of \textit{open edges}, meaning those edges that belong to a unique face, and let $\breve{V}$ be the set of \textit{open vertices}, where a vertex is defined to be open if it neighbors an open edge. The set of nonopen edges is denoted by $\mathring{E}\coloneqq E \setminus \breve{E}$ and the set of nonopen vertices is denoted by $\mathring{V}\coloneqq V \setminus \breve{V}$.\\

To prevent the algorithm from getting stuck before the whole forest is peeled away, each connected component of the erasure, referred to as a \textit{cluster erasure}, that contains a vertex $v \in \breve{V}$ is grown from $v$. Furthermore, during the growing process, edges are only added to the tree if they reach new vertices in $\mathring{V}$. This results in trees that contain at most one open vertex, located at their root. The whole spanning forest can therefore be peeled away without any problem.\\

\cref{alg:Erasure_decoder_surf_with_boundary} summarizes the decoding procedure for a cluster erasure of a general surface code $G = (V, E, F)$ with boundary. The full decoder consists of applying this procedure to each cluster erasure.

\begin{algorithm}[H]
    \caption{Erasure decoder for the surface code (with boundary)}\label{alg:Erasure_decoder_surf_with_boundary}
    \hspace*{\algorithmicindent} \textbf{Input\quad :} A surface $G=(V,E,F)$ with boundary, a cluster erasure $C \subset \varepsilon \subset \mathring{E}$, and the syndrome $\sigma \subset \mathring{V}$ of an $X$-error. \\
    \hspace*{\algorithmicindent} \textbf{Output\quad :} An $X$-error $P$ such that $P \subset \varepsilon$ and $\sigma(P) = \sigma$.
    \begin{algorithmic}[1]
        \State Initialize $A$ by $A = \emptyset$.
        \If{$C$ is connected to the boundary}
            \State Grow a spanning tree $F_C$ of $C$ from a vertex on the boundary.
        \EndIf
        \If{$C$ is not connected to the boundary}
            \State Grow a spanning tree $F_C$ of $C$ from any vertex.
        \EndIf
        \While{$F_C \neq \emptyset$}
            \State Pick a leaf edge $e=\{ u,v\}$ with pendant vertex $u \in \mathring{V}$ and remove $e$ from $F_C$.
            \If{$u \in \sigma$}
                \State Add $e$ to $A$, remove $u$ from $\sigma$, and flip $v$ in $\sigma$.
            \EndIf
            \If{$u \not\in \sigma$}
                \State Do nothing.
            \EndIf
        \EndWhile
        \State Return $P = \prod_{e \in A} X_e$.
    \end{algorithmic}
\end{algorithm}

It is known that \cref{alg:Erasure_decoder_surf_with_boundary} produces the most likely correction and that its time-complexity is linear in the number of qubits.

\begin{theorem}[Theorem 2 of \cite{delfosse2020linear}]\label{thm:erasure decoder surface}
For general surface codes with bounded degree and faces of bounded size, applying \cref{alg:Erasure_decoder_surf_with_boundary} to the graph and to its dual produces a linear-time maximum-likelihood decoder, i.e. a decoder that finds the most likely correction.
\end{theorem}


\section{Union-Find Decoder for the Surface Code}\label{sec:UF_surf}

The goal of the Union-Find decoder for the surface code \cite{delfosse2021almost} is to find an estimation of the $X$-error up to a stabilizer, given an erasure $\varepsilon$ and a syndrome $\sigma$. To achieve this, the decoder works in two stages. The first stage consists of converting the original erasure $\varepsilon$ and any possible Pauli errors into a modified erasure $\varepsilon'$. In other words, we look for a set of qubits that supports an error that is consistent with the syndrome. This stage is therefore called \textit{syndrome validation}. The second stage is then to simply apply the erasure decoder for the surface code.\\

Syndrome validation works by identifying all the clusters of erasures which are ``invalid'', and growing them until they become ``valid''. Similar to a cluster erasure, a \textit{cluster} is defined to be either a connected component of the erased qubits in the subgraph $(V, \varepsilon)$, or a single vertex outside the erasure supporting a syndrome bit. A cluster is said to be ``invalid'' or \textit{unsatisfiable} if it doesn't support an error that produces the syndrome on the cluster. A cluster that does support such an error is called \textit{satisfiable}. \\

For the surface code (without boundary), a cluster is satisfiable if and only if the cardinality of the syndrome on the cluster is even (Lemma 1 of \cite{delfosse2021almost}). Odd clusters must therefore be grown by adding edges until they connect with another odd cluster. The result of the merger of two odd clusters is an even cluster, which is correctable by the erasure decoder.\\

\cref{alg:UF_surf-naive} describes this growing and merging procedure.

\begin{algorithm}[H]
    \caption{Union-Find surface code decoder - Naive version}\label{alg:UF_surf-naive}
    \hspace*{\algorithmicindent} \textbf{Input\quad :} The set of erased positions $\varepsilon \subset E$ and the syndrome $\sigma$ of an error $E_X$. \\
    \hspace*{\algorithmicindent} \textbf{Output\quad :} An estimation $\mathcal{C}$ of $E_X$ up to a stabilizer.
    \begin{algorithmic}[1]
        \State Create the list of all odd clusters $C_1, \ldots, C_m$, and initialize the modified erasure $\varepsilon' = \varepsilon$.
        \While {there exists an odd cluster}
            \For {every odd cluster $C_i$}
                \State Grow $C_i$ by increasing its radius by one half-edge.
                \If {$C_i$ meets another cluster}
                    \State fuse and update parity.
                \EndIf
                \If {$C_i$ is even}
                    \State remove it from the odd cluster list.
                \EndIf
            \EndFor
        \EndWhile
        \State Add full edges that are in the grown clusters to $\varepsilon'$.
        \For {all even clusters}
            \State apply the erasure decoder for the surface code to find $\mathcal{C}$.
        \EndFor
    \end{algorithmic}
\end{algorithm}

\subsection{Surfaces with boundaries}

For surface codes with boundaries, a cluster is satisfiable if and only if it is an even cluster \textit{or} it connects to the boundary. The reason for this is that any remaining syndrome bit can always be canceled by an error coming from the boundary.\\

For surface codes with boundaries, we must therefore keep track of the \textit{boundary status} of each cluster, in addition to its \textit{parity}. We must also use the appropriate erasure decoder, capable of dealing with boundaries.\\

\cref{alg:UF_surf_boundary-naive} shows the modified procedure.

\begin{algorithm}[H]
    \caption{Union-Find surface code (with boundary) decoder - Naive version}\label{alg:UF_surf_boundary-naive}
    \hspace*{\algorithmicindent} \textbf{Input\quad :} The set of erased positions $\varepsilon \subset \mathring{E}$ and the syndrome $\sigma \subset \mathring{V}$ of an error $E_X$. \\
    \hspace*{\algorithmicindent} \textbf{Output\quad :} An estimation $\mathcal{C}$ of $E_X$ up to a stabilizer.
    \begin{algorithmic}[1]
        \State Create the list of all unsatisfiable clusters $C_1, \ldots, C_m$, and initialize the modified erasure $\varepsilon' = \varepsilon$.
        \While {there exists an unsatisfiable cluster}
            \For {all unsatisfiable clusters $C_i$}
                \State Grow $C_i$ by increasing its radius by one half-edge.
                \If {$C_i$ meets another cluster}
                    \State fuse and update satisfiability.
                \EndIf
                \If {$C_i$ meets the boundary}
                    \State update satisfiability.
                \EndIf
                \If {$C_i$ is satisfiable}
                    \State remove it from the unsatisfiable cluster list.
                \EndIf
            \EndFor
        \EndWhile
        \State Add full edges that are in the grown clusters to $\varepsilon'$.
        \For {all satisfiable clusters}
            \State apply the erasure decoder for the surface code to find $\mathcal{C}$.
        \EndFor
    \end{algorithmic}
\end{algorithm}

It has been shown that below the minimum distance $d$ of the surface code, the Union-Find decoder performs just as well as the most likely error (MLE) decoder.

\begin{theorem}[Theorem 1 of \cite{delfosse2021almost}]\label{thm:UF decoder surface}
If $t + 2s < d$, \cref{alg:UF_surf_boundary-naive} can correct any combination of $t$ erased qubits and $s$ $X$-errors.
\end{theorem}

\subsection{Almost-linear time-complexity}

\begin{algorithm}
    \caption{Union-Find surface code (with boundary) decoder - Almost linear time version}\label{alg:UF_surf_boundary-lin}
    \hspace*{\algorithmicindent} \textbf{Input\quad :} The set of erased positions $\varepsilon \subset \mathring{E}$ and the syndrome $\sigma \subset \mathring{V}$ of an error $E_X$. \\
    \hspace*{\algorithmicindent} \textbf{Output\quad :} An estimation $\mathcal{C}$ of $E_X$  up to a stabilizer.
    \begin{algorithmic}[1]
        \State Initialize cluster-trees, Support and boundary lists for all clusters.
        \State Create the list $\mathcal{L}$ of roots of unsatisfiable clusters.
        \While {$\mathcal{L} \neq \emptyset$}
            \State Initialize fusion list $\mathcal{F} = \emptyset$
            \For {all $u \in \mathcal{L}$}
                \State Grow the cluster $C_u$ a half-edge in the Table Support.
                \If {a new grown boundary edge $e$ is added in Support}
                    \State Update boundary at root $u$.
                \EndIf
                \If {a new grown interior edge $e$ is added in Support}
                    \State Add $e$ to the fusion list $\mathcal{F}$.
                \EndIf
            \EndFor
            \For {all $e = \{u,v\} \in \mathcal{F}$}
                \State Determine $\text{Find}(u)$ and $\text{Find}(v)$.
                \If {$\text{Find}(u)\neq \text{Find}(v)$}
                    \State Read the sizes at the roots.
                    \State Union the smaller cluster to the bigger cluster.
                    \State Append the boundary lists of the smaller cluster to the boundary list of the bigger cluster.
                \EndIf
                \If {$\text{Find}(u)=\text{Find}(v)$}
                    \State Remove $e$ from the list $\mathcal{F}$.
                \EndIf
            \EndFor
            \For {each $u \in \mathcal{L}$}
                \If {$\text{Find}(u) \neq u$}
                    \State remove $u$ from $\mathcal{L}$.
                \EndIf
            \EndFor
            \For {all $u \in \mathcal{L}$}
                \State Remove the vertices of the boundary lists of $u$ that are not boundary vertices.
            \EndFor
            \For {all $u \in \mathcal{L}$}
                \State Calculate the satisfiability of $C_u$.
                \If {$C_u$ is satisfiable}
                    \State Remove $u$ from $\mathcal{L}$.
                \EndIf
            \EndFor
        \EndWhile
        \State Erase all the edges that are fully grown in Support.
        \State Apply the erasure decoder for the surface code.
    \end{algorithmic}
\end{algorithm}

The decoder can be implemented in an almost-linear time $O(n\alpha (n))$, where $n$ is the number of qubits and $\alpha$ is the inverse of Ackermann's function. To achieve this almost-linear time-complexity, a Union-Find data-structure and algorithm \cite{galler1964improved,tarjan1975efficiency} are utilized, hence the name of the decoder.\\

Each cluster is represented by a tree, a \textit{cluster tree}, that consists of all the vertices\footnote{The vertices of a cluster include all the non-open vertices that neighbor at least one edge of the cluster.} of the cluster. An arbitrary vertex of the cluster is chosen as the root of the tree and serves as a representative of the cluster. To determine if two vertices belong to the same or to a different cluster, each vertex recursively looks at its parent until the root of the tree is found. The two vertices belong to the same cluster if and only if the roots are identical. Finding the root of a vertex is done with the \texttt{Find()} procedure.

The root of the cluster tree also stores the size of the cluster (the number of vertices of the cluster), its parity and its boundary status. The parity and boundary status are needed to determine the satisfiability of the cluster, while the size is needed to efficiently merge two clusters.

To merge two clusters, first the smallest cluster-tree is added as a subtree of the root of the largest cluster-tree, and then the size, parity and boundary status of the resulting tree are updated. Merging two clusters is done with the \texttt{Union()} procedure.\\

A look-up table, called \textit{Support}, is used to keep track of the status of the edges of the code. An edge is attributed both a \textit{filled status}: indicating whether it is \textit{Unoccupied}, \textit{Half-Grown} or \textit{Grown}, and a \textit{boundary status}: indicating whether it neighbors an open vertex or not.\\

To prevent having to update the whole Support table at each round of growth, which would lead to a quadratic time-complexity, a list of boundary vertices for each cluster is maintained. A \textit{boundary vertex} of a cluster is a vertex of the cluster for which at least one of the incident edges is not in the erasure $\varepsilon$. It is clear that an edge can only grow if it is incident to a boundary vertex. Only the filled status of these edges must therefore be updated at each round of growth.\\

\cref{alg:UF_surf_boundary-lin} describes the full almost-linear implementation of the Union-Find decoder for a surface code with boundary.



\bibliography{main_modified}

\end{document}